# Predicting Cricket Outcomes using Bayesian Priors

Mohammed Quazi, Joshua Clifford, Pavan Datta

## Abstract

This research develops a statistical modeling procedure to predict outcomes of future cricket tournaments. Proposed model provides an insight into the application of survey sampling to the team selection pattern by incorporating individual players' performance history based off of Bayesian priors not only against a particular opposition but also against any cricket playing nation - full member of International Cricket Council (ICC). A case study for the next ICC world cup 2023 is provided, and simulation results are discussed. Method is validated using 2020 Indian Premier League season. The method can predict probabilities of winning for each participating team.

## 1. Introduction

This research paper investigates the impact of incorporating sampling theory into team selection patterns and Bayesian priors to predict outcomes of limited overs, One-Day International (ODI) and Twenty20 (T20) cricket matches. The proposed idea could be extended to unlimited overs test cricket format as well. Often in cricket a particular player is most lethal only against a particular opponent. For instance, Australia's ace all-rounder Glenn Maxwell was dismissed in all four innings by Indian spinner Yuzvendra Chahal during Australia's tour of India in 2017. One-on-one battles define the outcome of a cricket game more often than not. So a predicting model that accounts for the impact of these one-on-one encounters is what is required in modern-day cricket, and this study aims to provide such a model, which could provide benefits such as helping team management make informed decisions regarding who gets to play and under what circumstances.

Cricket is a game involving a plethora of statistics: batsman's average score, batsman's highest score, bowler's economy rate, number of wickets taken by a bowler, average team score, and average runs saved by a fielder are just some examples. Previous research has investigated various methods of utilizing these statistics for modeling and making predictions. Swartz et. al (2009) developed a discrete probabilistic model to simulate outcomes. Modeling match outcomes in annual sporting contests, including cricket, using likelihood functions is studied by Baker and Scarf (2006). Kimber and Hansford (1993) discussed a nonparametric approach to estimate batting averages in cricket, though this study makes use of the traditional averages reported by the International Cricket Council (ICC). Preston and Thomas (2002) have studied probabilities of victory in cricket by shedding light on interruptions due to bad weather. Most recently, Pathak and Wadhwa (2016) have tried to predict the outcome of ODI cricket matches using modern classification techniques such as support vector machines and random forests. Interested readers are referred to Allsopp and Clarke (2004) as well.

The next section provides a short background of cricket. Section 3 discusses the proposed model. Section 4 presents the model's predictions for the ICC 2023 Cricket World Cup (CWC) followed by a validation of the model in its application to the 2020 Indian Premier League (IPL) season. The final section summarizes the results and addresses the scope of the model.

# 2. The 'Gentleman's Game' and its Laws

Cricket is predominantly played in three different formats: ODI, T20, and Test. ODI is a 50-over game and usually takes about six hours to finish. T20 is a 20-over game that usually takes about three hours, is the shortest format, and is considered the most entertaining and a high revenue generating format. Test cricket is played over five days with unlimited overs in a game. Test cricket is the pioneering format of all cricket but has lost its popularity over the years. The ICC has repeatedly taken steps, such as introducing a World Test Championship, to revive test cricket.

Cricket originated in England and gained its popularity through English colonies in the Indian subcontinent, Australasia, parts of Africa, and West Indies. Played initially only by the Englishmen, it started being referred to as the 'gentleman's game.' Cricket provided the natives with an attractive opportunity to beat the English at their own game. Outside of England, it has gained the largest fan following in the Indian subcontinent and Australia. It is the second most popular sport in the world, with two and a half billion fans, second only to soccer's four billion fans (worldatlas.com, 2020).

A game of cricket is played between two teams of eleven players. These eleven players have different roles in a team. For example, a cricket team could consist of four bowlers, four batsmen, two all-rounders, and one wicket-keeper. ODI and T20 games consist of two innings per game (one for each team), whereas a test match has four innings (two for each team). A cricket game of any format is won by the team that scores at least one run more than the opposition, irrespective of which team bats first. The rules of cricket are very complicated, and owing to the complications, its rules are rightly referred to as 'laws of the game.'

The ICC ODI CWC takes place every four years, whereas an ICC T20 International World Cup occurs biennially. England won the last ODI CWC, hosted by England and Wales, in 2019. India is considered the heavy favorite to win the next tournament in 2023, in part due to the advantage of being the host nation, and the fact that India won in 2011 as a tournament co-host.

Most recently, T20 has gained a lot of popularity and interest among the fans. The Indian Premier League (IPL), which was founded in 2008, is the most revenue generating and popular of all the T20 leagues as well as the most well-attended cricket league (smh.com.au, 2020). The Big Bash League in Australia, Pakistan Super League in Pakistan, Caribbean Premier League in West Indies, and Mzansi Super League in South Africa are some of the other premier T20 league competitions.

# 3. Methodology

The proposed predictive model makes use of stratified random sampling, which is used widely in sampling theory. To understand stratified random sampling, one must first be familiar with simple random sampling (SRS), the most basic form of sampling. In SRS, all the elements in the population have an equal probability of being in the sample, and any subset has equal probability of being selected as the sample. For example, if we need to select a sample of four students from a population of ten students, each subset ($\mathcal{S}$) of four students has probability

$$P(\mathcal{S}) = \frac{1}{\binom{10}{4}}$$

and each student has 4/10 probability of being in the sample. In stratified random sampling, the population of interest is divided into subcategories, called strata, such that each element belongs only to one stratum, and SRS is applied independently to each stratum. Effective stratification ensures the elements in each stratum are more similar to one another compared to elements from other stratum, which often results in more precise estimates of population parameters. While this sampling design is fairly straightforward, survey sampling can become very complicated. Zhang et. al (2019) and Valliant (1987) have worked on estimation in more complex survey designs resulting from use of stratified random sampling. Lohr (2010) is a great resource to learn about different survey sampling techniques.

Player roles form different strata for this study. Strata are composed of different types of players based on their role in the team, including fast bowler, spinner, all-rounder-fast bowler, and batsman. The particular strata used for the two competitions this method is applied on will be described in the next section. A stratified random sample of size eleven is selected to simulate team selection for a match. Next, each selected player's statistics against a particular opponent are obtained, specifically past batting average and highest score against that opponent. These two quantities are used to derive parameters for a gamma distribution, which is used to produce a random number representing the predicted score for that player against the particular opposition. The gamma probability density function is

$$f(x|\alpha,\beta) = \frac{1}{\Gamma(\alpha)} x^{\alpha-1} e^{-x/\beta}, \quad 0 \le x < \infty, \quad \alpha, \beta > 0 \tag{1}$$

with

$$\mathrm{E}(X) = \alpha\beta \quad \mathrm{Var}(X) = \alpha\beta^2$$

where $\mathrm{E}(X)$ is the expected value of the random variable $X$, which can be thought of as the average of the probability distribution. $\mathrm{Var}(X)$ is the variance of the random variable $X$, random variable in this study is the runs scored by a player in a given game. $\Gamma(\alpha)$ denotes the gamma function evaluated at $\alpha$.

There are two parameters to derive: $\alpha$ and $\beta$. We use a player's average score against the opponent as an estimate of $\mathrm{E}(X)$ and generate 50,000 potential candidates for $\beta$ from 0.01 to 5,000. We then assume that every player has at most a 5% chance to surpass their highest score against a particular team every time he gets an opportunity to bat against them. These steps lead to the simultaneous equations

$$\alpha = \frac{\text{Average score}}{\beta} \tag{2}$$

$$P(X > \text{Highest score}) \le 0.05 \tag{3}$$

that are solved to estimate $\alpha$ and $\beta$. For example, New Zealand's captain, Kane Williamson, averages 54.61 runs and has a highest score of 118 against England, which leads to estimated parameters of $\alpha$ = 86.68 and $\beta$ = 0.63 for him against England. For each simulated matchup against England,

Williamson's runs scored are randomly sampled from a gamma distribution with these parameters. See Figures 5 and 6 in the Appendix for example estimated gamma distributions of predicted runs scored for selected players in various matchups for the CWC (including Williamson) and a player from the IPL. Predicted scores using this method for all the players on a team in a matchup are added to get the team predicted score that is compared against the opponent team's score to obtain the winner of that simulated matchup. This process is repeated 10,000 times for the full simulation results to best account for any variation due to chance.

As mentioned in the last section, the team that scores at least one run more than the opposition wins the game in cricket, irrespective of which team gets to bat first. The team batting first is decided by a coin toss at the start of the game. The winner of the toss gets to choose to bat first or put opposition to bat first. Our approach makes the prediction independent of the coin toss results, thereby providing more simplicity to model and management for making decisions on team selection. Preston and Thomas (2000) have made an effort to study differences between strategies for batting first and second. Going back to the example of Chahal against Maxwell, clearly the presence of Chahal in the team thwarts Maxwell's ability to score runs, and our model accounts for these one-on-one matchups. The approach of deriving the parameters of a gamma distribution for this study could be thought of as a Bayesian statistical approach to parameter estimation where the prior distribution is heavily influenced by historical data, see Christensen et. al (2011). Also, a gamma distribution is ideal for prediction here because of its nonnegative support and asymptotic property, its ability to approach normal distribution. Since runs scored in cricket are always integer values, the fact that the gamma distribution is a continuous distribution can be dealt with by just rounding the randomly sampled real values (of double or float precision) to their nearest integer estimates.

Finally, the proposed model accounts for debutant's performances and the effect of playing conditions as well. Examples are discussed in the next section.

# 4. Results

In this section, results obtained after applying the proposed model to ICC Men's ODI 2023 CWC are discussed. Next, the model is validated by using it to predict outcomes for the IPL T20 2020 season. Specific considerations for each competition format are also discussed in each subsection.

## 4.1. ICC Men's ODI 2023 CWC

First, the stratified random sampling design employed for this tournament will be explained. The six strata formed by player roles are fast bowler, spinner, all-rounder-fast bowler, all-rounder-spinner, batsman, and wicket-keeper. Figure 1 depicts the sampling design. Historically, spinners have heavily impacted games played in the Indian subcontinent. The high temperatures in the region tend to result in a dry pitch with wider cracks than pitches in more moderate climates. Spin bowlers, or spinners, can take advantage of the irregularities of the pitch to deceive the batsmen with their bowling delivery. Since the tournament is going to be hosted by India, we can expect that spinners will be heavily relied upon during the games. In contrast, Australian and English conditions favor fast bowlers over spinners. For this reason, allotting spinners into their own stratum is reasonable.

Web scraped data collected spans over 11,000 individual ODI innings that cover every competitive game played among the full members of the ICC from January 1999 until June 2020 (espncricinfo.com, 2020a). One hundred and ninety-five international players are considered for this particular case study. The tournament will feature ten teams that will play against each other in a league format at the group stage. The winner of each game gets two points and the losing side gets zero points, and a draw results in one point for each side. Therefore, a given team can have a maximum point tally of eighteen points from the nine games in group stage. The top four teams at the end of the group stage qualify for the semis, where the first placed team plays the fourth placed team and the second placed team plays against the third placed team. This exact format was used in the 2019 edition of the tournament as well.

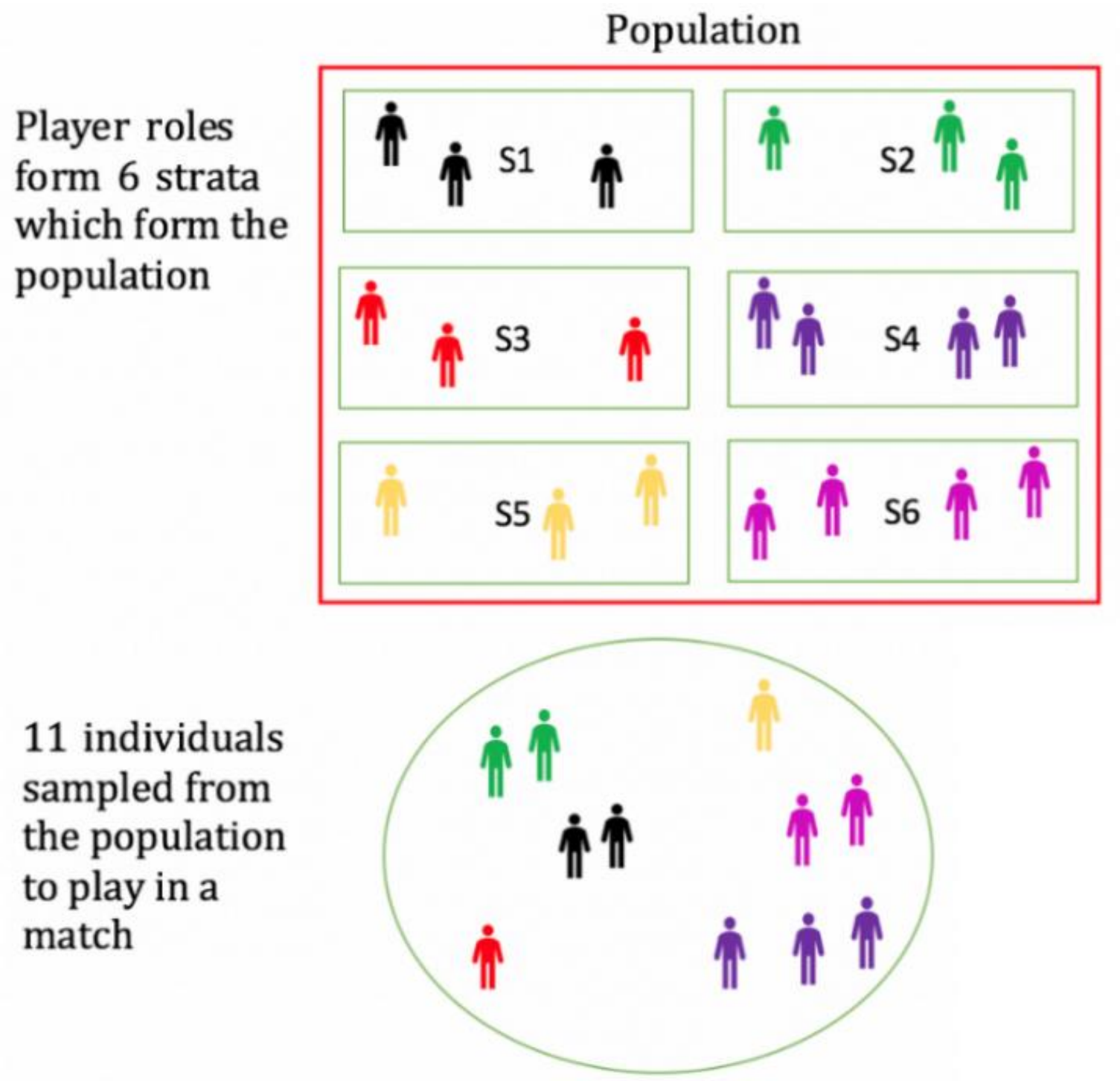

*Figure 1: Sampling scheme employed in this study. S1-S6 are strata. The oval region shows individuals sampled using the stratified random sampling method.*

Only India has already qualified for the ICC 2023 CWC because of its host status. For the purpose of this study, twelve teams have been included so that no team (full member of ICC) that may compete in the CWC is omitted from the analysis. For more information about the qualifying process, refer to (icc-cricket.com) and (espncricinfo.com, 2020b). Predicted head-to-head winning probability between these twelve teams is shown in Table 1, and the actual head-to-head record for the time frame of the scraped data is shown in Table 2 (espncricinfo.com, 2020c). See Figure 5 (Appendix) for estimated gamma distributions of predicted runs for selected players in their potential matchups at the ICC 2023 CWC.

*Table 1: Predicted win/loss ratio (in %) at the ICC 2023 CWC.*

| | Losing Team | | | | | | | | | | | |
|---|---|---|---|---|---|---|---|---|---|---|---|---|
| **Winning Team** | Afghanistan | Australia | Bangladesh | England | India | Ireland | New Zealand | Pakistan | South Africa | Sri Lanka | West Indies | Zimbabwe |
| Afghanistan | | 0.17 | 42.5 | 3.75 | 0.03 | 94.9 | 1.04 | 30.6 | 0.00 | 7.56 | 19.5 | 52.2 |
| Australia | 99.8 | | 86.5 | 59.6 | 64.9 | 100 | 99.6 | 27.8 | 0.95 | 8.02 | 86.7 | 100 |
| Bangladesh | 56.8 | 13.3 | | 2.84 | 2.59 | 100 | 0.99 | 12.1 | 0.40 | 2.59 | 95.4 | 100 |
| England | 96.1 | 39.6 | 97.0 | | 10.5 | 93.2 | 69.4 | 38.4 | 12.0 | 34.7 | 35.7 | 88.1 |
| India | 99.9 | 34.5 | 97.2 | 89.1 | | 100 | 99.9 | 68.5 | 40.2 | 99.9 | 99.3 | 100 |
| Ireland | 4.65 | 0.00 | 0.00 | 6.54 | 0.00 | | 0.00 | 0.00 | 0.00 | 32.1 | 0.00 | 40.7 |
| New Zealand | 98.9 | 0.38 | 98.9 | 30.0 | 0.01 | 100 | | 84.6 | 19.8 | 100 | 55.5 | 89.4 |
| Pakistan | 68.6 | 71.0 | 87.6 | 60.9 | 30.8 | 100 | 14.5 | | 90.0 | 100 | 78.9 | 100 |
| South Africa | 100 | 99.0 | 99.5 | 87.7 | 58.4 | 100 | 79.5 | 9.73 | | 100 | 71.1 | 100 |
| Sri Lanka | 91.9 | 91.6 | 97.2 | 64.6 | 0.13 | 67.1 | 0.00 | 0.00 | 0.00 | | 70.6 | 99.5 |
| West Indies | 80.0 | 12.9 | 4.34 | 63.4 | 0.66 | 100 | 44.1 | 20.6 | 28.4 | 28.4 | | 96.9 |
| Zimbabwe | 46.6 | 0.00 | 0.00 | 11.2 | 0.00 | 58.3 | 10.1 | 0.00 | 0.00 | 0.45 | 2.91 | |

*Table 2: Actual win/loss ratio for national teams (in %) since January 1999 until June 2020.*

| | Losing Team | | | | | | | | | | | |
|---|---|---|---|---|---|---|---|---|---|---|---|---|
| **Winning Team** | Afghanistan | Australia | Bangladesh | England | India | Ireland | New Zealand | Pakistan | South Africa | Sri Lanka | West Indies | Zimbabwe |
| Afghanistan | | 0.00 | 37.5 | 0.00 | 0.00 | 48.1 | 0.00 | 0.00 | 0.00 | 25.0 | 33.3 | 60.0 |
| Australia | 100 | | 90.0 | 59.8 | 57.0 | 80.0 | 63.1 | 78.2 | 45.5 | 62.9 | 73.2 | 89.5 |
| Bangladesh | 62.5 | 5.00 | | 19.0 | 17.2 | 70.0 | 29.4 | 15.6 | 19.0 | 16.3 | 39.5 | 64.4 |
| England | 100 | 37.0 | 81.0 | | 34.3 | 76.9 | 40.9 | 57.4 | 47.8 | 44.3 | 61.4 | 83.3 |
| India | 83.3 | 34.9 | 79.3 | 59.7 | | 100 | 46.9 | 49.2 | 47.4 | 59.8 | 59.2 | 84.6 |
| Ireland | 50.0 | 0.00 | 20.0 | 15.4 | 0.00 | | 0.00 | 14.3 | 0.00 | 0.00 | 8.33 | 46.2 |
| New Zealand | 100 | 29.2 | 70.6 | 52.3 | 46.9 | 100 | | 50.8 | 33.9 | 42.4 | 58.5 | 66.7 |
| Pakistan | 100 | 20.0 | 84.4 | 38.3 | 50.8 | 71.4 | 45.8 | | 36.2 | 54.8 | 68.0 | 88.4 |
| South Africa | 100 | 48.5 | 81.0 | 41.3 | 49.1 | 100 | 57.6 | 62.1 | | 58.1 | 74.5 | 94.1 |
| Sri Lanka | 75.0 | 33.9 | 79.1 | 50.8 | 33.3 | 100 | 47.5 | 41.7 | 40.3 | | 67.7 | 78.6 |
| West Indies | 62.5 | 19.6 | 55.3 | 29.5 | 34.2 | 83.3 | 29.3 | 30.0 | 19.6 | 25.8 | | 72.1 |
| Zimbabwe | 40.0 | 5.26 | 35.6 | 12.5 | 15.4 | 46.2 | 28.6 | 4.65 | 5.88 | 16.7 | 23.3 | |

Figure 2 shows a comparison of the two tables using a plot. This comparison provides insight into certain teams the simulation method favors in various matchups and when that favoritism differs from expectations based on results of recent team matchups. For instance, Sri Lanka has performed fairly well in matchups against India, New Zealand, Pakistan, and South Africa, but the simulation views Sri Lanka's chances of victory against these teams in the 2023 CWC as extremely low. India is favored more by the simulation than would be expected from recent results in its matchups against most of the countries in contention. Similarly, South Africa is also favored more by the simulation in numerous matchups, with the exception of matchups against Pakistan. The results of this comparison only indicate the simulation predictions are different in some cases than what one would predict only looking at results of recent matchups. However, if the simulation method's predictive performance can be validated (see next section), then the argument can be made that the simulation is effectively incorporating additional useful information by considering player performance in these recent matchups.

Figure 3 shows the predicted team standings at the end of the group stage, based on generating 10,000 simulated samples. Results indicate that India has the highest chance of qualifying for the semifinals and a 24% chance of winning the cup. Nevertheless, somewhat unexpectedly, Pakistan is the favorite, with a 47% chance of winning the tournament if it makes it to the semifinals. Such a performance would cement Pakistan's 'unpredictable' tag and set up a replay of the ICC Champions

Trophy Final in 2017, which Pakistan won despite being underdogs against India. South Africa, a historic underachiever at the CWC, is projected to shed its reputation for choking and qualify for the semifinals, with a 21% chance of winning the tournament. Five-time CWC winner Australia has an 8% chance of winning the cup. To incorporate the playing conditions into the predictions, we considered an increase in probability of including spinners in the team, thereby slightly reducing probability of including fast pace bowlers. To account for debutants' performances, we considered domestic cricket, first class cricket, international reserve team cricket, and international under-19 cricket performances for players' predicted runs.

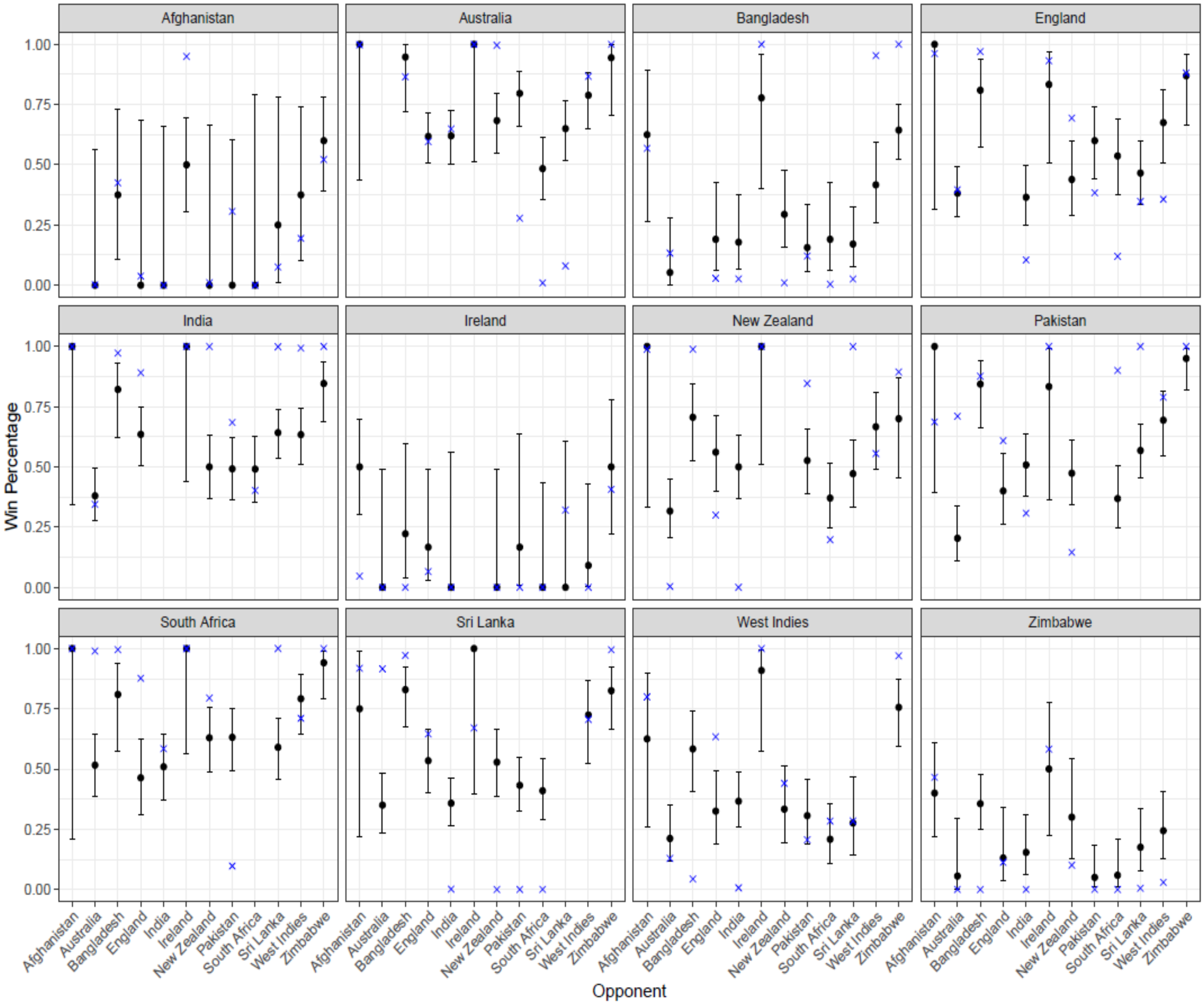

*Figure 2: Comparison of recent matchup win % versus simulation predictions. Black dots represent actual win % with a 95% confidence interval. Blue X's represent simulation predicted win %.*

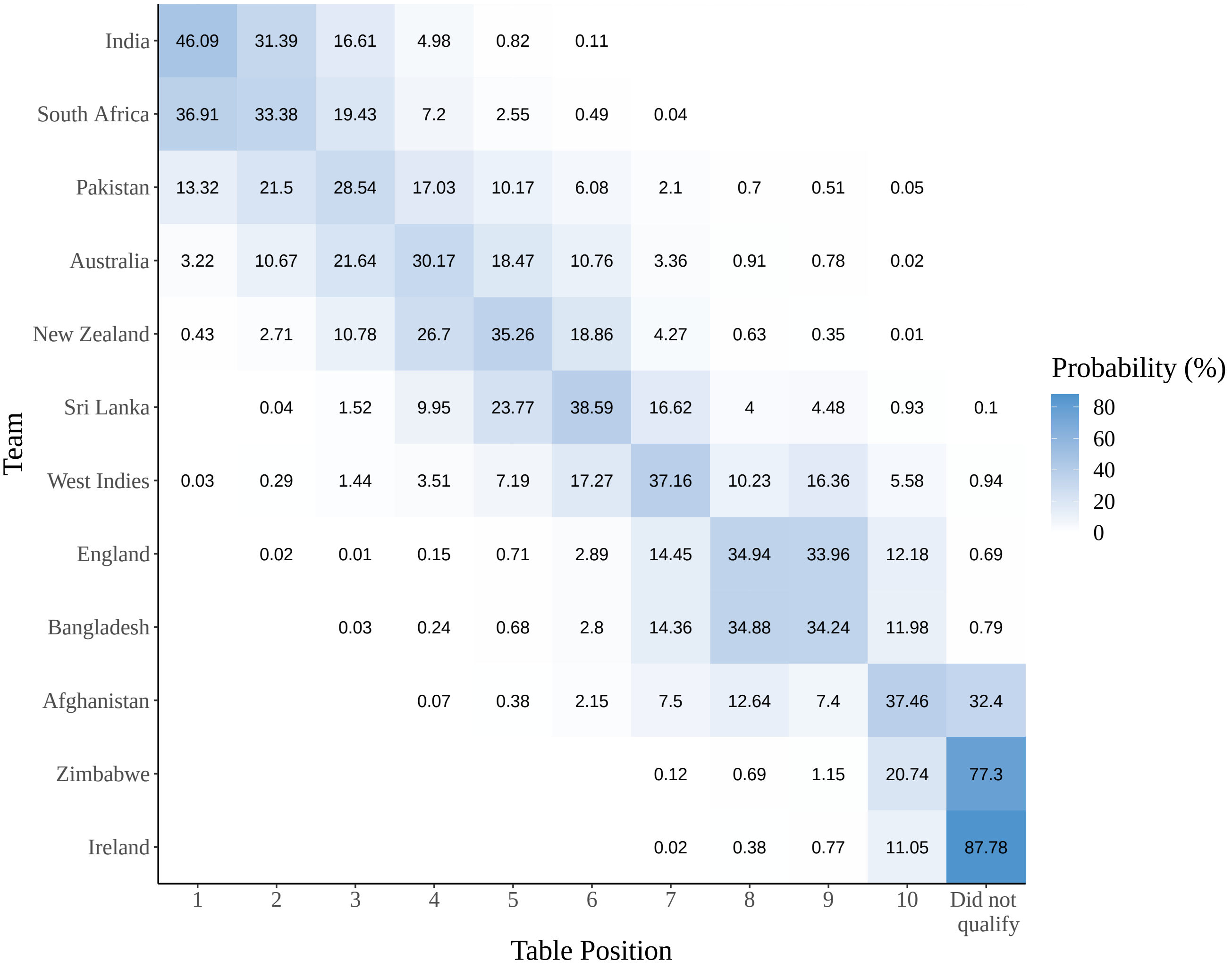

*Figure 3: ICC 2023 CWC predicted team standings probability distribution (in %).*

## 4.2. Model Validation: IPL 2020 Application

We validate the proposed model in this subsection by applying it to predict outcomes for the IPL 2020 season. Data for over 130 players' profiles across eight teams who participated in the IPL 2020 season was web scraped (cricimetric.com, 2020). This data covers every game played from the first season of IPL in 2008 to the IPL 2019 edition. Due to the restriction that a starting lineup cannot have more than four overseas players (from any country other than India) in the IPL, the overseas players were assigned into their own stratum, out of which only four players are selected and the remaining seven are from other strata. This rule was implemented to promote local Indian talent. The retained players by the teams were stratified as described in the previous section.

In IPL each team plays against every other team twice in the league stage. A win gains two points and loss zero points for the team. Draws result in a super over, which means the winner is eventually decided after two or more short innings per team. The top four teams at the end of this stage qualify for the playoffs. The maximum points possible at the end of the league stage for any team is twenty-eight points. Ties on the table are decided by net run rate. In the first playoff, the first-place team plays against the second-place team, and the winner proceeds to the final. In the eliminator playoff, the third and fourth placed teams are pitted against each other, and the winner qualifies for the second-playoff to play the loser of the first playoff. The winner of the second playoff proceeds to the final game of the season. This process gives an advantage to the two best performing teams at the league stage.

The simulation method was applied, and the predicted team standings (along with the actual final IPL 2020 standings) can be seen in Figure 4. The proposed model has predicted the first three positions with 100% accuracy and correctly predicted four of the top five teams. Additionally, no team ended in a position outside of the realm of possibility according to the simulation.

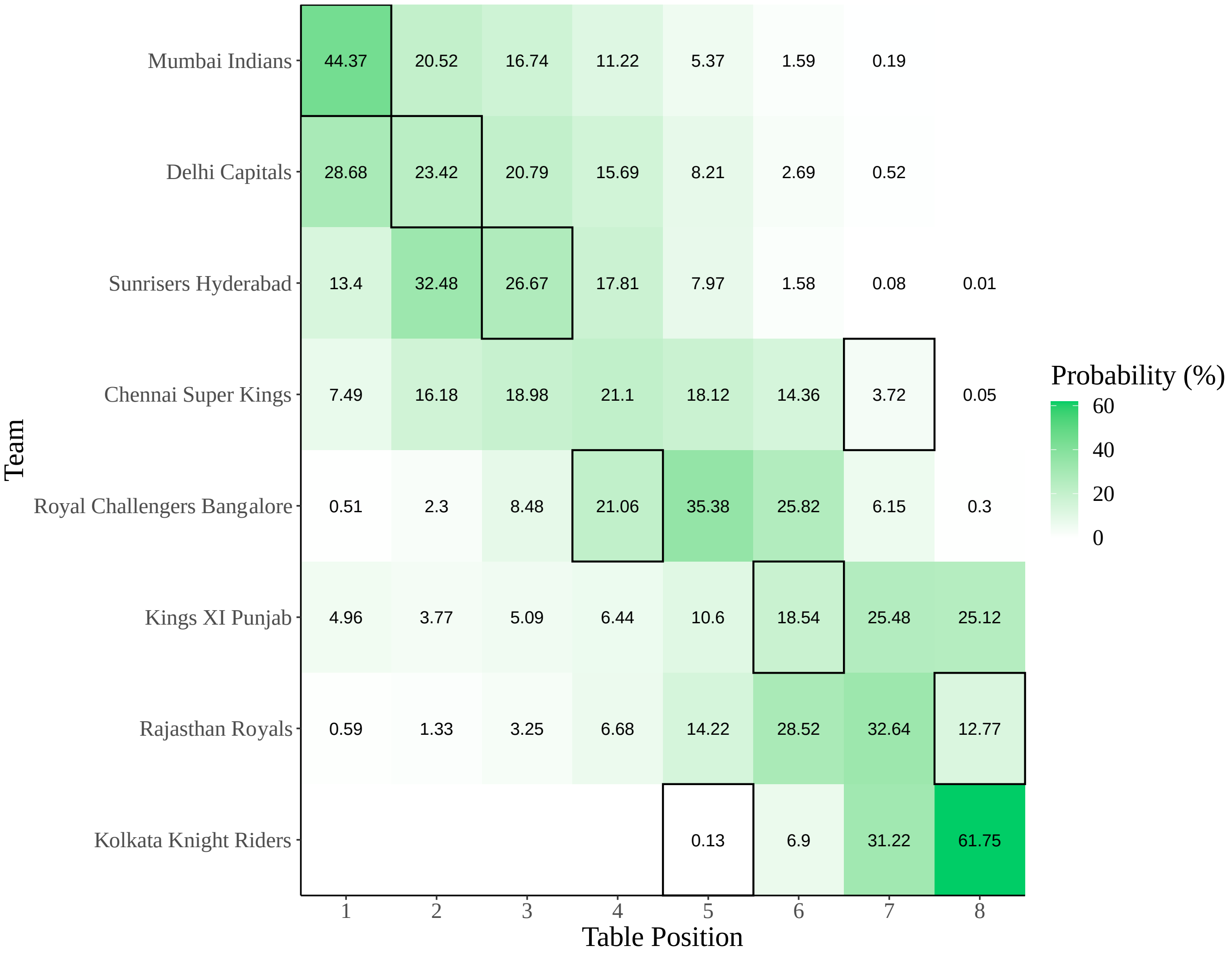

*Figure 4: IPL 2020 predicted team standings probability distribution (in %). Actual IPL 2020 standings are shown for each team with a black rectangle.*

The most notable differences between the predictions and actual standings are the Chennai Super Kings underperforming against the simulation's expectations and the Kolkata Knight Riders overperforming. Chennai has historically been among the two most successful teams in IPL based on performance, but we believe the reason for underperforming this season was the unavailability of Suresh Raina, a batsman who had performed consistently well for them in the past. Also, Dwayne

Bravo's injury meant that he could play only 6 games out of 14. On the other hand, Kolkata made Eoin Morgan its captain in the middle of this season, which made the whole team step up its game. The probability of this final table in Figure 4 based on the probability distributions provided by the simulation is the highest among all possible outcomes. Overall, the simulation performed well in predicting the final IPL 2020 team standings. See Figure 6 (Appendix) for estimated gamma distributions of predicted runs for a player's matchups in IPL 2020.

From the standings in Figure 4 and predicted probability of winning in Table 3, the method gives Sunrisers Hyderabad and Chennai Super Kings together less than a 1% chance of winning the IPL 2020 championship. The Delhi Capitals have a chance of about 27%, while the Mumbai Indians have a 73% chance of winning the tournament. As predicted by the model to be the most likely outcome, the Mumbai Indians turned out to be the actual winners of IPL 2020 as well, winning against the Delhi Capitals in the final game.

*Table 3: Predicted win/loss ratio (in %) for IPL 2020.*

| | **Losing Team** | | | | | | | |
|---|---|---|---|---|---|---|---|---|
| **Winning Team** | Chennai Super Kings | Delhi Capitals | Kings XI Punjab | Kolkata Knight Riders | Mumbai Indians | Rajasthan Royals | Royal Challengers Bangalore | Sunrisers Hyderabad |
| Chennai Super Kings | | 98.5 | 95.7 | 100 | 2.84 | 45.2 | 72.2 | 4.37 |
| Delhi Capitals | 1.28 | | 70.2 | 99.9 | 26.3 | 98.6 | 93.8 | 97.8 |
| Kings XI Punjab | 3.88 | 29.0 | | 49.8 | 57.5 | 14.2 | 48.0 | 28.6 |
| Kolkata Knight Riders | 0.00 | 0.03 | 49.1 | | 0.00 | 0.09 | 38.5 | 0.00 |
| Mumbai Indians | 96.9 | 72.2 | 40.8 | 100 | | 2.80 | 99.1 | 93.0 |
| Rajasthan Royals | 53.4 | 1.36 | 84.8 | 99.8 | 96.9 | | 3.91 | 0.72 |
| Royal Challengers Bangalore | 26.8 | 5.88 | 51.0 | 60.8 | 0.79 | 95.6 | | 0.00 |
| Sunrisers Hyderabad | 95.5 | 2.12 | 69.7 | 100 | 6.69 | 99.1 | 100 | |

# 5. Conclusion and Impact on the Game

The proposed model provides an alternative to existing techniques of predicting cricket game outcomes by using the ideas of sampling theory and statistical distributions. While the model depends on historical data, it accounts for the playing conditions as well as those players with no prior history in the competition. The proposed model shows promise for effectively predicting results of games and competitions, as proved by the model validation. It not only provides above par prediction accuracy but also flexibility to apply to other cricket tournaments.

The method can predict probabilities of winning, which is of interest to fans and also could inform setting gambling odds. The method can also be adapted to recommend team selection strategies to inform teams' approaches to increase chances of success in a game. This study could be applied to all league-format cricket tournaments, including the super sixes format, a format which also goes by Hong Kong cricket sixes. The model could be even implemented for other sports as well, such as to predict soccer outcomes by switching average number of runs with average goals scored per game. We address the potential applications of the model for cricket in the following subsections.

### 5.1. Pregame Team Strategy

Use of the model could provide an opportunity for team management to simulate different team subsets and evaluate predicted performance so that an optimal set of eleven players are picked to play the game. It provides bowlers with an insight into the opposition's batting lineup, which will help reduce the number of runs conceded by bowlers. It could also help to predict the opposition's starting lineup well before the actual teams are announced just before the game, thereby providing a team with a potential advantage over its opponent.

### 5.2. In-Game Team Strategy

Gamma parameters can be generated with more information as the game progresses for a particular player while he is batting, thus making the model a dynamic one. If the predicted score is high, then this will provide the bowlers with an objective to attack a particular batsman. On the other hand, merely having players like Chahal on the side will not help, but he can be available to bowl when players like Maxwell come out to bat. So the model can help the captain of a side decide how and when to allot the bowling overs among his bowlers.

### 5.3. Postgame Strategy

At the end of the game, gamma parameters can be generated again, and scores for the next game can be predicted, providing the team a head start in planning for the next game. Total predicted scores can help a team decide before the game to choose to bat first or second if it wins the coin toss. For instance, if the predicted team score is too high, then the team may prefer to bat second because they would have to chase a relatively lower score. In that case, the team could decide to rest players in the second inning to avoid injuries.

### 5.4. Betting

The most basic application of the model to gambling would be to use the predicted probability for a team winning a matchup from the simulation to calculate odds and set betting lines. A more realistic application would likely be incorporating the results of the simulation to inform the more complex models in use to establish betting lines today. Further research could be done into how to best tune the simulation methodology toward this purpose.

### 5.5. Debutant's Performance

The model framework as described lets us incorporate previous performances of players like Joshua Phillip and Devdutt Padikkal, who made their IPL debuts last year (in the 2020 season). Although there is no data available for them from the 2008-2019 IPL editions, their other league performances were effective in producing meaningful IPL 2020 predictions.

Data repository: https://github.com/mquazi/cricket_2020

# References

Allsopp, P. E. and Clarke, S. R. (2004). Rating teams and analysing outcomes in one-day and test cricket. *Journal of the Royal Statistical Society: Series A (Statistics in Society)*, *167*(4), 657-667.

Baker, R. and Scarf, P. (2006). Predicting the outcomes of annual sporting contests. *Journal of the Royal Statistical Society: Series C (Applied Statistics)*, *55*(2), 225-239.

Christensen, R., Johnson, W., Branscum, A. and Hanson, T. E. (2011). *Bayesian ideas and data analysis: an introduction for scientists and statisticians*. CRC Press.

cricmetric.com (2020). Retrieved from https://www.cricmetric.com/tools/

espncricinfo.com (2020a). Retrieved from https://stats.espncricinfo.com/ci/engine/stats/index.html

espncricinfo.com (2020b). Retrieved from https://www.espncricinfo.com/story/_/id/27384084/how-teams-secure-qualification-rank-no-1-32

espncricinfo.com (2020c). Retrieved from https://stats.espncricinfo.com/ci/engine/stats/index.html?class=2;...

icc-cricket.com (2020a). Retrieved from https://www.icc-cricket.com/media-releases/1310135

icc-cricket.com (2020b). Retrieved from https://www.icc-cricket.com/rankings/mens/player-rankings/odi/batting

Kimber, A. C. and Hansford, A. R. (1993). A statistical analysis of batting in cricket. *Journal of the Royal Statistical Society. Series A (Statistics in Society)*, *156*(3), 443-455.

Lohr, S. L. (2010). *Sampling design and analysis, 2nd edition*. Chapman & Hall/CRC Press.

Pathak, N. and Wadhwa, H. (2016). Applications of modern classification techniques to predict the outcome of ODI cricket. *Procedia Computer Science*, *87*, 55-60.

Preston, I. and Thomas, J. (2000). Batting strategy in limited overs cricket. *Journal of the Royal Statistical Society: Series D (The Statistician)*, *49*(1), 95-106.

Preston, I. and Thomas, J. (2002). Rain rules for limited overs cricket and probabilities of victory. *Journal of the Royal Statistical Society: Series D (The Statistician)*, *51*(2), 189-202.

smh.com.au (2020). Retrieved from https://www.smh.com.au/sport/cricket/big-bash-league-jumps-into-top-10-of-most-attended-sports-leagues-in-the-world-20160110-gm2w8z.html

Swartz, T. B., Gill, P. S. and Muthukumarana, S. (2009). Modelling and simulation for one-day cricket. *Canadian Journal of Statistics*, *37*(2), 143-160.

Valliant, R. (1987). Generalized variance functions in stratified two-stage sampling. *Journal of the American Statistical Association*, *82*(398), 499-508.

worldatlas.com (2020). Retrieved from https://www.worldatlas.com/articles/what-are-the-most-popular-sports-in-the-world.html

Zhang, G., Cheng, Y. and Lu, Y. (2019). Generalised variance functions for longitudinal survey data. *Statistical Theory and Related Fields*, *3*(2), 150-157.

# Appendix

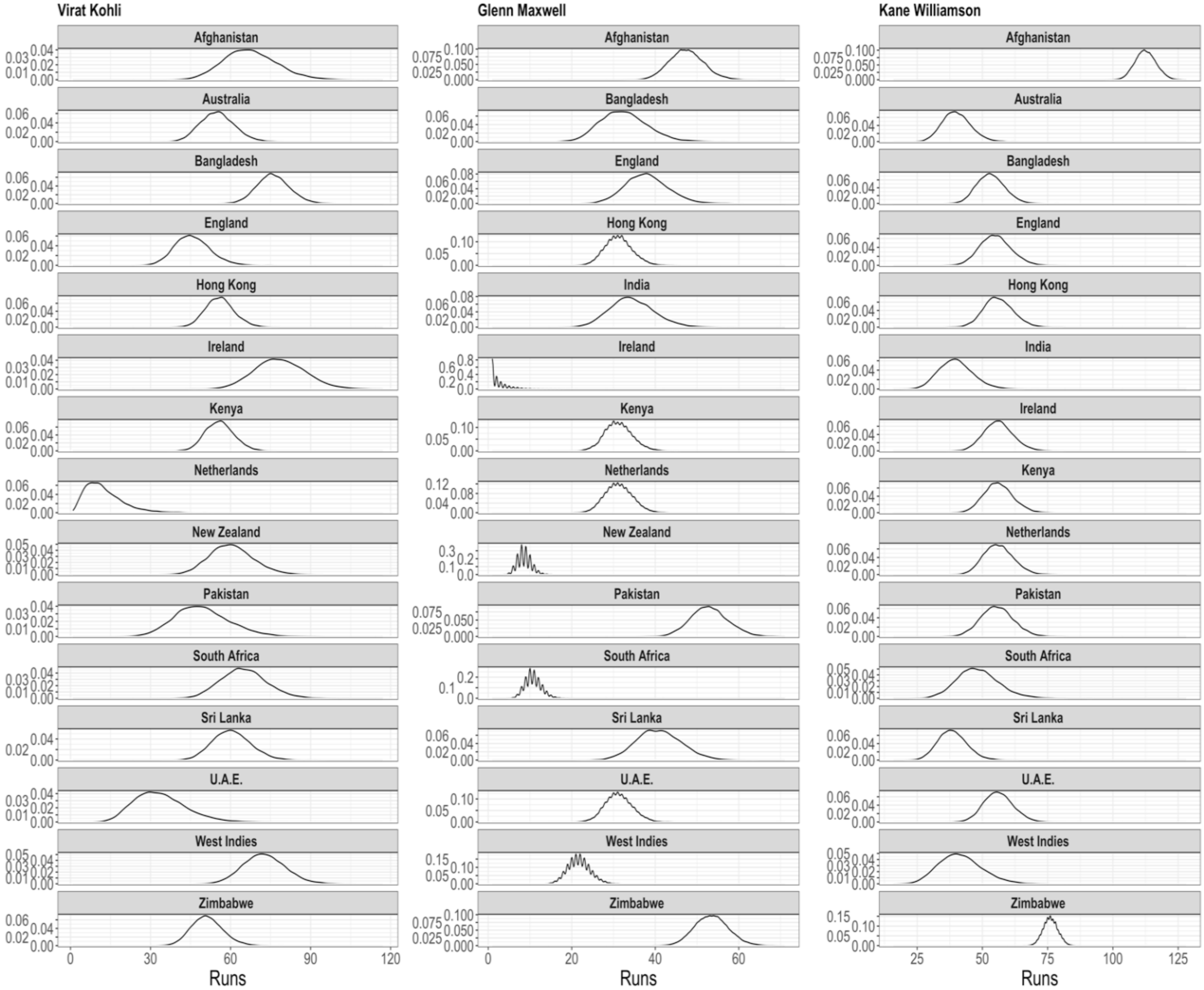

*Figure 5: Probability distributions from which the predicted scores are generated for India's batsman Virat Kohli, Australia's all-rounder Glenn Maxwell, and New Zealand's batsman Kane Williamson. Notice the range of horizontal axes is different for the three players. These plots are consistent with the fact that Kohli is currently ranked number 1 while Kane is at 8 and Maxwell at 20 in the ICC ODI batting rankings (icc-cricket.com, 2020b) because, though each player has certain matchups they are expected to perform particularly better or worse than usual, Kohli is generally expected to score the most runs in various matchups (50-60), followed by Williamson (45-55), then Maxwell (30-40).*

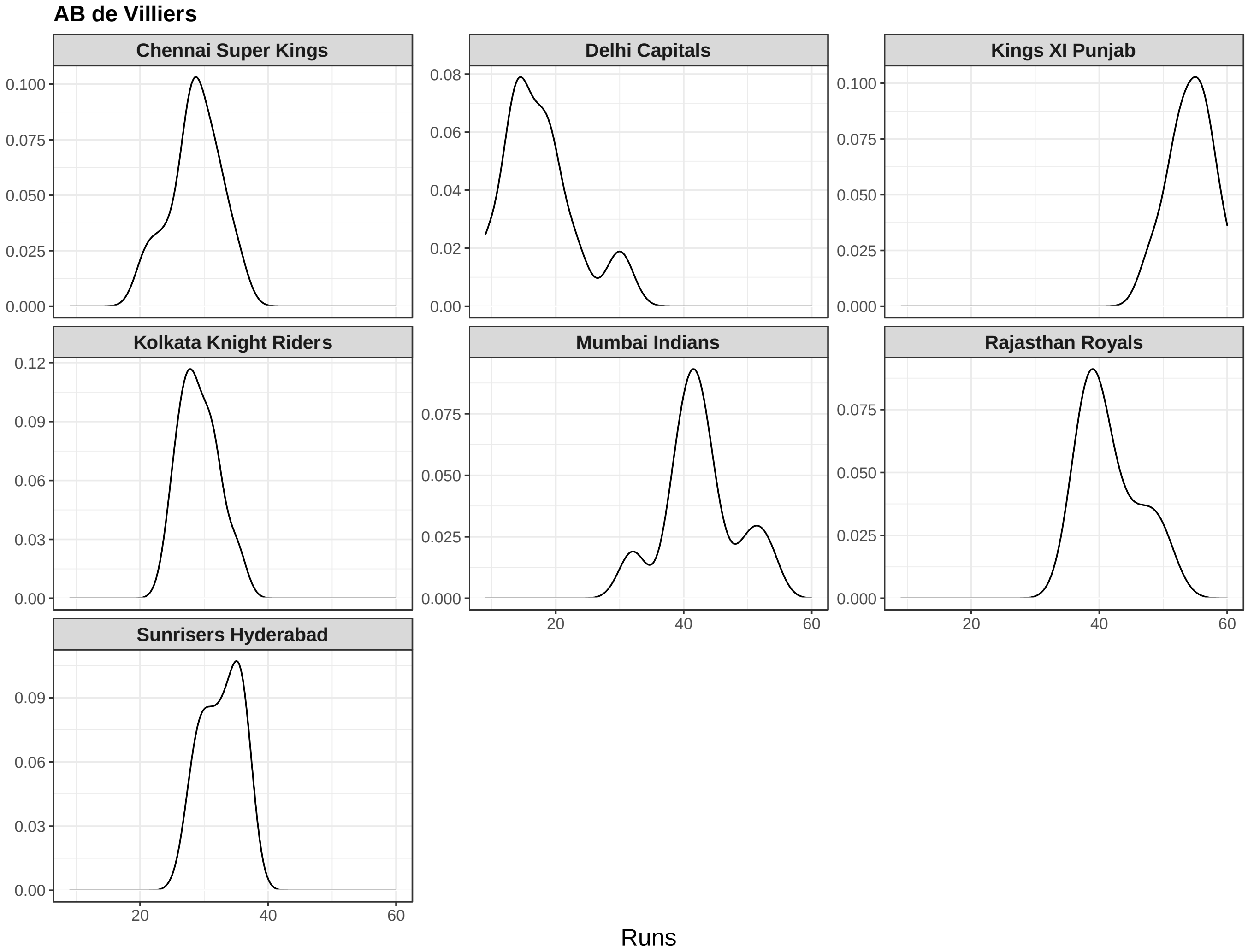

*Figure 6*: *Probability distributions from which the predicted scores for IPL 2020 are generated for Royal Challengers Bangalore's wicket-keeper batsman AB de Villiers. Based on the distributions, he was expected to score the most runs against Kings XI Punjab and the least against the Delhi Capitals. As expected from the simulation, he actually averaged only 22 against Delhi Capitals in IPL 2020 season, which is his third lowest average this season. Most curves look normally distributed because of gamma distribution's asymptotic property.*